\documentclass[11pt]{article}
\usepackage[margin=1in]{geometry}
\usepackage{graphicx}

\usepackage{amssymb}
\usepackage{url}
\usepackage{bm}
\usepackage{graphicx}
\usepackage{float}
\usepackage{floatflt}
\usepackage{slashed}

\def\beq{\begin{equation}}
\def\eeq#1{\label{#1}\end{equation}}
\def\eeqn{\end{equation}}

\def\beqa{\begin{eqnarray}}
\def\eeqa#1{\label{#1}\end{eqnarray}}
\def\eeqan{\end{eqnarray}}

\let\bar=\overbar

\def\Dslash{\not{\hbox{\kern-4pt $D$}}}
\def\dslash{\not{\hbox{\kern-2pt $\del$}}}

\def\msb{{\bar{\ssstyle M \kern -1pt S}}}

\def\Title#1{\begin{center} {\Large {\bf #1} } \end{center}}
\def\Author#1{\begin{center} {\normalsize {\sc #1} } \end{center}}
\def\Institution#1{\begin{center} {\normalsize {\it #1} } \end{center}}
\def\Abstract#1{\noindent {\normalsize {\bf Abstract:} {\normalfont #1}}}
\def\Conference{\vspace{4mm}\begin{raggedright} {\normalsize {\it Talk presented at the 2019 Meeting of the Division of Particles and Fields of the American Physical Society (DPF2019), July 29--August 2, 2019, Northeastern University, Boston, C1907293.} } \end{raggedright}\vspace{4mm}}

\begin{document}

\Title{Gauge couplings in a multicomponent dark matter scenario}

\Author{Reagan Thornberry, Alejandro Arroyo, Caden LaFontaine, Gabriel Frohaug, \\ Dylan Blend, and Roland E Allen}

\Institution{Department of Physics and Astronomy, Texas A\&M University, College Station, Texas 77843, USA}

\Abstract{We consider the gauge couplings of a new dark matter candidate and find that they are comparable to those of a neutralino.}

\Conference 

In earlier papers we introduced a new dark matter candidate with favorable
characteristics~\cite{DM1,DM2,DM3,DM4}, which we have called a Higgson and represented by $H^{0}$.
This is the lowest-mass of a set of particles generically labeled $H$ to
distinguish them from Higgs bosons $h$ and the Higgsinos $\widetilde{h}$ of
supersymmetry. A Higgs field $\phi $ is interpreted as an amplitude mode in
an extended structure with multicomponent fields $\Phi $, somewhat in
analogy with the Higgs/amplitude modes of superconductors~\cite{Varma}:
\begin{eqnarray}
\Phi =\phi \chi 
\label{eq1}
\end{eqnarray}
or%
\begin{eqnarray}
\Phi ^{r}=\phi ^{r}\chi ^{r}
\label{eq2}
\end{eqnarray}
where $\phi $ is a gauge multiplet, consisting of complex scalar fields $%
\phi ^{r}$, and $\chi ^{r}$ is a  multicomponent spinor with $\chi ^{r\dag
}\chi ^{r}=1$. 
Summations are never implied over a repeated gauge index $r$, but are always
implied over repeated coordinate indices like $\mu ,\nu =0,1,2,3$ and $%
k=1,2,3$, as well as the index $j$ 
labeling gauge generators. A standard notation is used below, with Pauli
matrices represented by $\sigma ^{k}$, gauge covariant derivatives having
the convention $D_{\mu }=\partial _{\mu }-iA_{\mu }$, and the vector potential and field strength
tensor having the forms
\begin{eqnarray}
A_{\mu } &=&A_{\mu }^{j}t^{j}\quad ,\quad F_{\mu \nu }=F_{\mu \nu
}^{j}t^{j}\quad  \\
F_{\mu \nu }^{j} &=&\partial _{\mu }A_{\nu }^{j}-\partial _{\nu }A_{\mu
}^{j}+c_{ii^{\prime }}^{j}A_{\mu }^{i}A_{\nu }^{i^{\prime } }\; .
\label{eq3}
\end{eqnarray}%
The $S^{\mu \nu }=\frac{1}{2}\sigma ^{\mu \nu }$ below are the Lorentz generators which act on Dirac spinors. 
Natural units $\hbar=c=1$ are used, with the $(-+++)$ convention for the metric tensor.

In earlier papers~\cite{DM3,DM4} we emphasized the couplings of $H^{0}$ via the
exchange of Higgs bosons $h^{0}$, since these can lead to substantial
cross-sections for spin-independent scattering off nuclei, and may therefore
be the most promising mechanism for direct detection of dark matter WIMPs.
We found that naturalness suggests that the Higgs-mediated couplings of $H^{0}
$ are comparable to those for a neutralino consisting of an optimal admixture
of Higgsinos and neutral gauginos. 

Here we turn to the gauge couplings and
find a similar result for coupling via the $Z^{0}$.  Of course, the Higgs-mediated interaction
emphasized earlier is still the primary mechanism for spin-independent direct detection
of both $H^{0}$~\cite{DM3,DM4} and the lowest-mass 
neutralino $\chi ^{0}$~\cite{Baer-Tata,susy-DM-1996}. (In the
present multicomponent scenario, both $H^{0}$ and $\chi ^{0}$ are stable
dark matter WIMPs, with $H^{0}$ having a mass which is rigorously bounded as 
$\leq 125$ GeV/$c^{2}$ and $\chi ^{0}$ presumably having a substantially
higher mass.)

We also generalize the treatment to an arbitrary Higgson field $H$
which may be neutral or charged. Its gauge couplings are contained in two
terms: 
\begin{eqnarray}
S_{H}=S_{1}+S_{1}^{\prime }
\label{eq6}
\end{eqnarray}
\begin{eqnarray}
S_{1}=\int d^{4}x\,H^{\dag }\left( x\right) D^{\mu }D_{\mu
}H\left( x\right) 
\label{eq7}
\end{eqnarray}%
\begin{eqnarray}
S_{1}^{\prime }=\int d^{4}x\,\left( \frac{1}{2}H^{\dag }\left(
x\right) \,S^{\mu \nu }F_{\mu \nu }\,H\left( x\right) +h.c.\right) 
\label{eq8}
\end{eqnarray}%
where the notation is defined above. $H\left( x\right) $ is an excitation of
a bosonic multicomponent field $\Phi \left( x\right) =\Phi _{0}+H\left(
x\right) $, where $\Phi _{0}$ may contain a condensate. 

If we now treat $%
\Phi \left( x\right) $ and $H\left( x\right) $ as quantum fields with
canonical quantization, $\Phi $ is required
to satisfy the usual equal-time commutation relations, with the momentum
density $\Pi \left( x\right) =\partial _{0}\Phi ^{\dag }\left( x\right) $
obtained from the Lagrangian of (\ref{eq7}). 
$\Phi $ and $H $ must also be written in terms of the
usual destruction and creation operators $c_{p\kappa }$ and $d_{p\kappa }^{\dag }$, , where $\kappa $ and
$p$ label particle and antiparticle states $\sqrt{2p_{0}}c_{\kappa p}^{\dag }\left\vert 0\right\rangle $ 
and $\sqrt{2p_{0}}d_{\kappa p}^{\dag }\left\vert 0\right\rangle $. These
requirements imply that
\begin{eqnarray}
H\left( x\right) =\int \frac{d^{3}p}{\left( 2\pi \right) ^{3}}\frac{1}{\sqrt{%
2p_{0}}}\sum\limits_{s}\left( c_{p}^{s}\,U_{p}^{s}\,e^{ip\cdot
x}+d_{p}^{s\,\dag }V_{p}^{s}\,e^{-ip\cdot x}\right) 
\label{eq9}
\end{eqnarray}%
with  $p_0$ on-shell and 
\begin{eqnarray}
\left[ H\left( \overrightarrow{x},x^{0}\right) ,\partial _{0}H^{\dag
}\left( \overrightarrow{x}^{\prime },x^{0}\right) \right] _{-}=\left[ \Phi
\left( \overrightarrow{x},x^{0}\right) ,\Pi \left( \overrightarrow{x}%
^{\prime },x^{0}\right) \right] _{-}=i \mathbf{1}\delta \left( 
\overrightarrow{x}-\overrightarrow{x}^{\prime }\right) 
\label{eq10}
\end{eqnarray}%
where $\mathbf{1}$ represents the identity matrix and%
\begin{eqnarray}
\left[ c_{p}^{s},c_{p^{\prime }}^{s^{\prime }\,\dag }\right] _{-}=\delta
^{ss^{\prime }}\left( 2\pi \right) ^{3}\delta \left( \overrightarrow{p}-%
\overrightarrow{p}^{\prime }\right) \quad ,\quad \left[ d_{p}^{s},d_{p^{%
\prime }}^{s^{\prime }\,\dag }\right] _{-}=\delta ^{ss^{\prime }}\left( 2\pi
\right) ^{3}\delta \left( \overrightarrow{p}-\overrightarrow{p}^{\prime
}\right) \;,
\label{eq11}
\end{eqnarray}%
with the other commutators equaling zero. Here $\overrightarrow{p}$ is the
3-momentum and $\overrightarrow{x}$ the 3-space coordinate. This is
satisfied if%
\begin{eqnarray}
\sum\limits_{s}U_{p}^{s}\,U_{p}^{s\,\dag }=\mathbf{1}\quad
,\quad \sum\limits_{s}V_{p}^{s}\,V_{p}^{s\,\dag }=\mathbf{1}
\label{eq12}
\end{eqnarray}%
since  
\begin{eqnarray*}
\left[ H\left( \overrightarrow{x},x^{0}\right) ,\partial _{0}H^{\dag
}\left( \overrightarrow{x}^{\prime },x^{0}\right) \right] _{-} &=&\int \frac{%
d^{3}p}{\left( 2\pi \right) ^{3}}e^{i\overrightarrow{p}\cdot \left( \overrightarrow{x}-\overrightarrow{x}^{\prime }\right)
}\left( +ip_{0}\right) \frac{1}{2p_{0}}\sum\limits_{s}\left(
U_{p}^{s}\,U_{p}^{s\,\dag }+V_{p}^{s}\,V_{p}^{s\,\dag }\right)  \\
&=&i\mathbf{1}\delta \left( \overrightarrow{x}-\overrightarrow{x}%
^{\prime }\right) \;.
\label{eq13}
\end{eqnarray*}

For unpolarized cross-sections involving a vertex with $H$
and $\overline{H}$ (now representing particles in external
lines), the Feynman diagram rules will give a contribution with sums
involving $u_{p}^{s}\,$, $u_{p}^{s\,\dag }$, $v_{p}^{s}\,$, $v_{p}^{s\,\dag }
$, so the external lines will give a dimensionless contribution of order 1.
But the contribution from (\ref{eq7}) of the vertex itself is of order $p_{0 }$ -- the particle energy--  and the contribution  from (\ref{eq8}) of its vertex is of order $\left | \overrightarrow{P} \right |$
 -- the magnitude of the 3-momentum of the gauge boson $Z^{0}$.  

For spin 1/2 fermions with mass $M$, on the other hand, the corresponding
normalization is 
\begin{eqnarray}
\sum\limits_{s}u_{p}^{s}\,\overline{u}_{p}^{s}=\slashed{p}+M\quad ,\quad
\sum\limits_{s}v_{p}^{s}\,\overline{v}_{p}^{s}=\slashed{p}-M
\label{eq14}
\end{eqnarray}%
and the external lines will make a contribution of order $M$. But the
contribution of the vertex itself is a dimensionless coupling constant of
order 1.

In the following discussion, we will neglect second-order processes like that in the right-hand panel of Fig.~\ref{fig2}. We will also 
assume that the particle $H$ has small 3-momentum 
$\overrightarrow{p}$, as is the case for slow-moving dark matter particles in
the present universe (with speeds $\sim 10^{-3}c$) and near threshold in
other situations. The contribution of (\ref{eq8}) will therefore be neglected. (Since the incident and scattered particle have small 3-momenta 
$\overrightarrow{p}$ and $\overrightarrow{p}^{\prime }$, the same is true of
the emitted $Z^{0}$ with 3-momentum  $\overrightarrow{P}=\overrightarrow{p}-\overrightarrow{p}^{\prime }$. )
\begin{figure}
\centering
\includegraphics[width=0.27\textwidth]{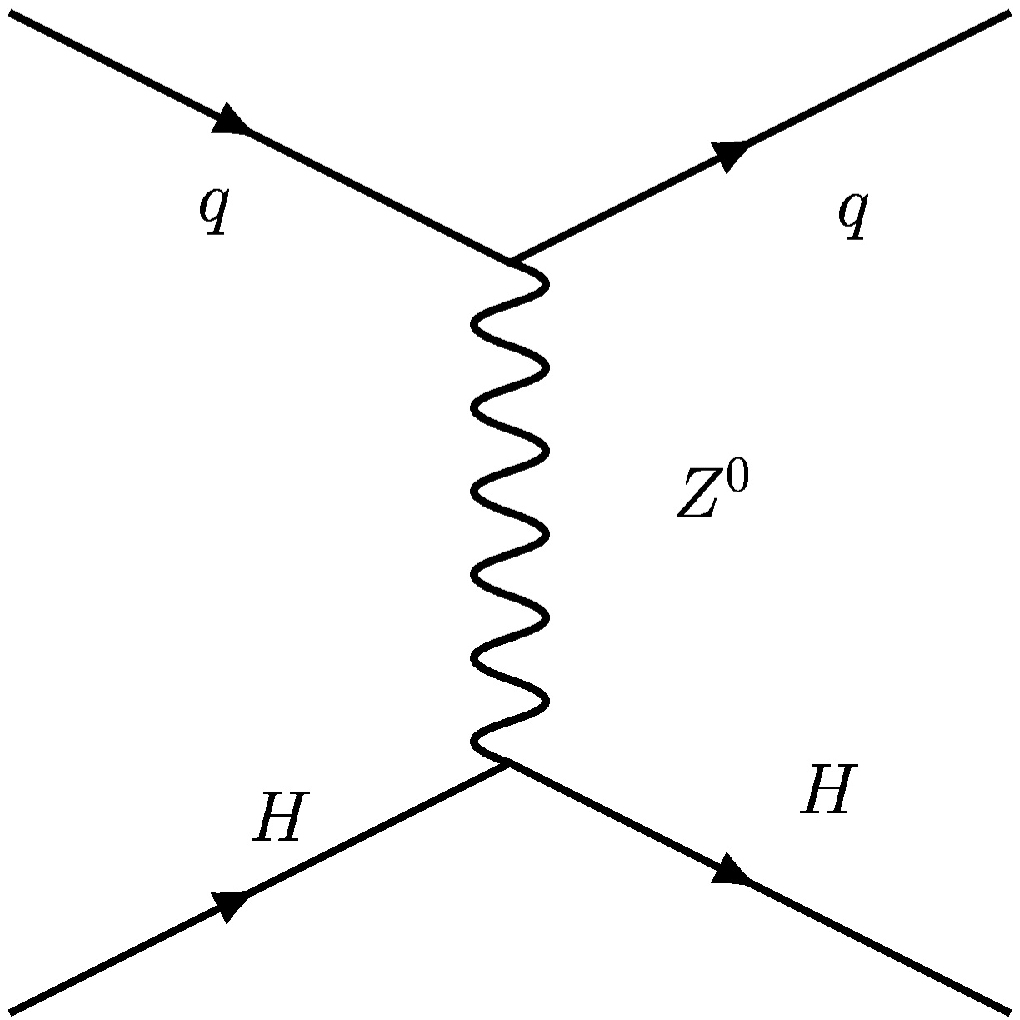} 
\hspace*{80pt}
\includegraphics[width=0.37\textwidth]{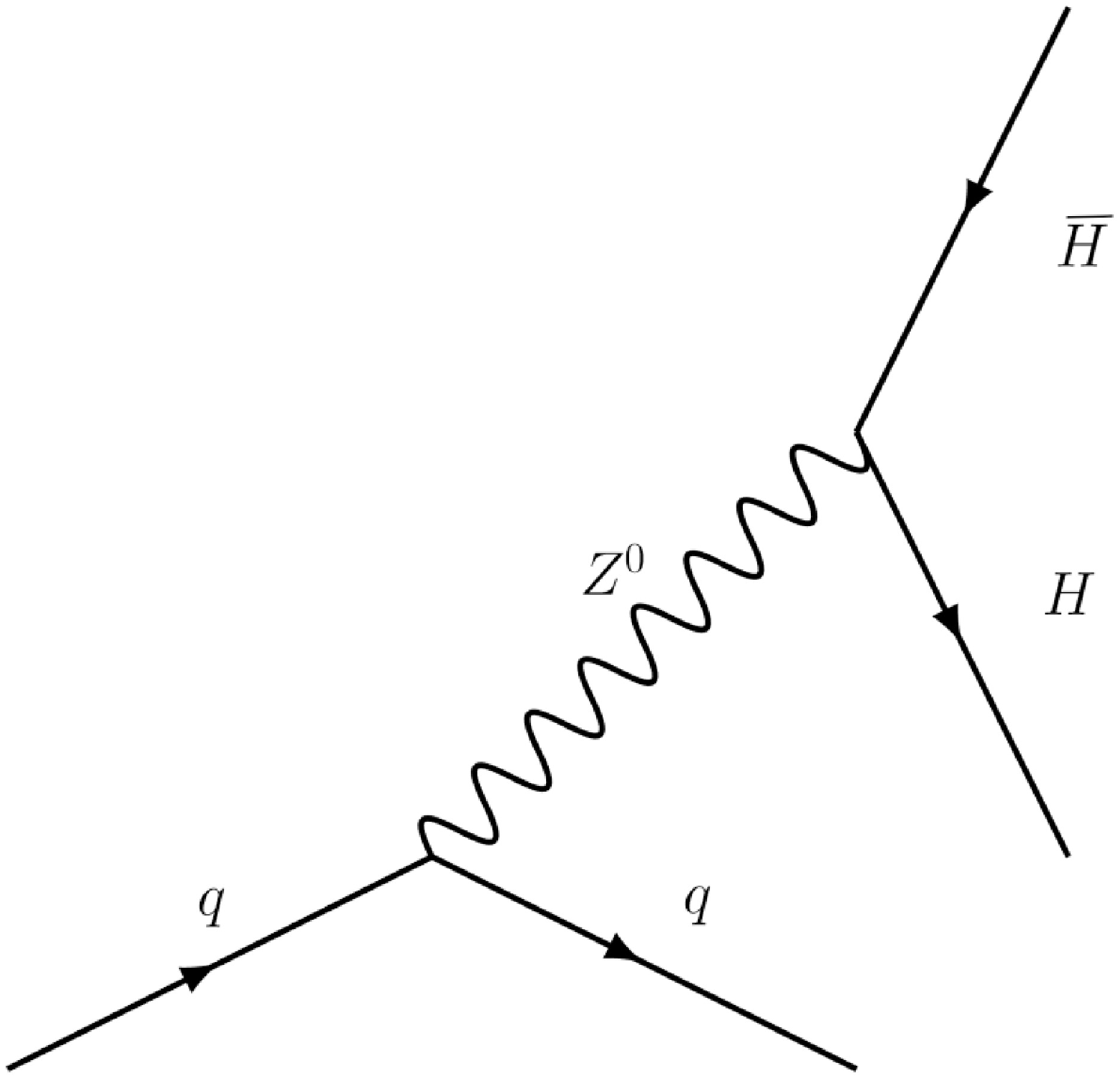} 
\caption{Left: Scattering of a Higgson $H$ off a quark in a nucleus via $Z^0$ exchange. This process is relevant to direct detection. Right: One mechanism for creation of a $H$ $\overline{H}$ pair. This process is relevant to creation in a collider or the early universe.}
\label{fig1}
\end{figure}
\begin{figure}
\centering
\includegraphics[width=0.45\textwidth]{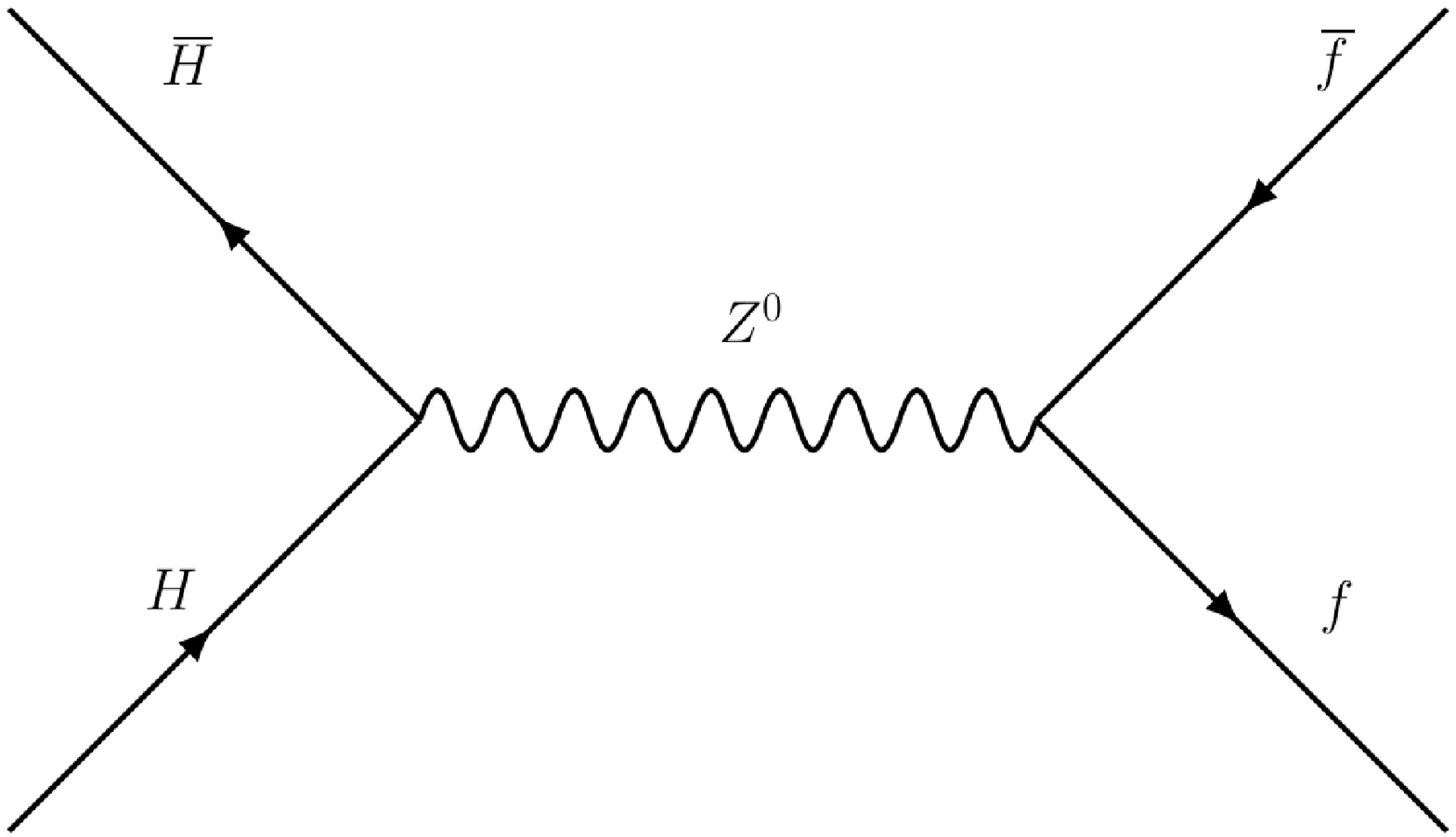} 
\hspace*{80pt}
\includegraphics[width=0.27\textwidth]{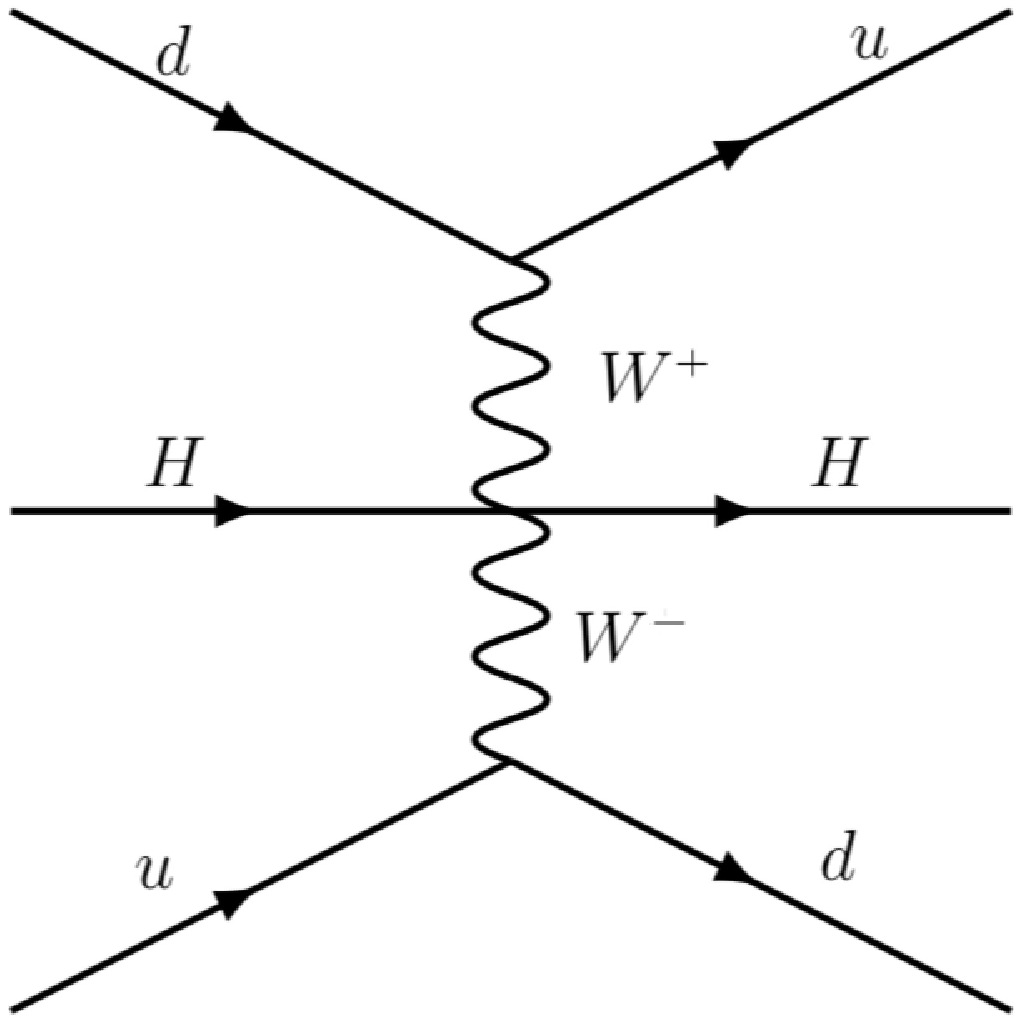} 
\caption{Left: Annihilation of $H$ $\overline{H}$ pair into Standard Model fermions via coupling to $Z^0$. The subsequent processes follow just as for the neutralino. Right: A second-order process not considered in the present treatment because the cross-section is small.}
\label{fig2}
\end{figure}

The other term $S_{1}$ of (\ref{eq7}), on the other hand, makes a
contribution to the vertex which is given by the 4-momentum $p$\ of the
particle, and for a slow-moving dark matter particle $p_{0}$ is
approximately equal to the rest mass energy $m$ (in natural units). It
follows that the contribution of the gauge couplings of a Higgson $H$ via $%
Z^{0}$ exchange is comparable to that of a neutralino $\chi $, although
somewhat weaker if the Higgson mass $m$ is much less than the neutralino
mass $M$. One should also remember that $Z^{0}$ exchange gives weaker
cross-sections than Higgs exchange for direct detection.

Also, for indirect detection, the annihilation process in the left-hand panel of Fig.~\ref{fig2} involves
particles which each have $p_0\approx m$, so that there is again a large gauge contribution from (\ref{eq7}). 

Similarly, if a pair of particles is produced in a collider, via the process
in the right-hand panel of Fig.~\ref{fig1}, each particle
emerges with $p_0 \geq  m$, yielding a large value for the gauge contribution to the cross-section. And the same conclusion holds when
particles are created in the early universe through this mechanism. 

In an earlier paper~\cite{DM3} we demonstrated the existence of Higgs-mediated 
interactions for $H^0$, and then emphasized them because (1) they 
will produce larger (spin-independent) cross-sections for direct detection 
and (2) the gauge interactions of Higgsons will be  somewhat weaker
than those of neutralinos if the ratio $m/M$ of Higgson mass to neutralino
mass is small. 

The main conclusion of this paper, however, is that the gauge
interactions of Higgsons $H$ -- including the new dark matter candidate $%
H^{0}$\ -- are comparable to those of neutralinos, including the lowest-mass $%
\chi ^{0}$, which is the other stable WIMP in the present multicomponent scenario.

\end{document}